\documentclass{aa}  
\usepackage[]{natbib}
\bibpunct{(}{)}{;}{a}{}{,}
\usepackage{graphicx}
\usepackage{txfonts}

\usepackage[]{hyperref}

\newcommand{\water}{H$_2$O}

\newcommand{\methane}{CH$_4$}
\newcommand{\kms}{km~s$^{-1}$}
\newcommand{\kp}{$K_\mathrm{P}$}

\newcommand{\pname}{HD~189\,733~b}
\newcommand{\sname}{HD~189\,733}
\newcommand{\micron}{$\mu$m}

\begin{document} 

\title{Exoplanet atmospheres with GIANO}

\subtitle{I. Water in the transmission spectrum of \pname}

\author{M. Brogi\inst{1,2} \and P. Giacobbe\inst{3} \and G. Guilluy\inst{3,4} \and R. J. de Kok\inst{5,6,7} \and A. Sozzetti\inst{3} \and L. Mancini\inst{8, 9, 1} \and A. S. Bonomo\inst{3} }

\institute{Department of Physics, University of Warwick, Coventry CV4 7AL, UK \\
              \email{m.brogi@warwick.ac.uk} \and
Center for Astrophysics and Space Astronomy (CASA), University of Colorado Boulder, Boulder, CO 80309, USA \and
INAF-Osservatorio Astrofisico di Torino, via Osservatorio 20, 10025, Pino Torinese, Italy \and
Dipartimento di Fisica, Universit\`a di Torino, via P. Giuria 1, I-10125 Torino, Italy \and
SRON Netherlands Institute for Space Research, Sorbonnelaan 2, 3584 CA Utrecht, The Netherlands  \and
Leiden Observatory, Leiden University, Postbus 9513, 2300 RA, Leiden, The Netherlands \and
Utrecht University, Department of Physical Geography, PO Box 80115, 3508 TC, Utrecht, The Netherlands \and
Dipartimento di Fisica, Universit\`a di Roma Tor Vergata, via della Ricerca Scientifica 1, 00133 Roma, Italy \and
Max Planck Institute for Astronomy, K\"{o}nigstuhl 17, 69117 Heidelberg, Germany
}

   \date{}

 
  \abstract
{High-resolution spectroscopy ($R\ge$ 20\,000) at near-infrared wavelengths can be used to investigate the composition, structure, and circulation patterns of exoplanet atmospheres. However, up to now it has been the exclusive dominion of the biggest telescope facilities on the ground, due to the large amount of photons necessary to measure a signal in high-dispersion spectra.}
{Here we show that spectrographs with a novel design - in particular a large spectral range - can open exoplanet characterisation to smaller telescope facilities too. We aim to demonstrate the concept on a series of spectra of the exoplanet \pname\ taken at the Telescopio Nazionale Galileo with the near-infrared spectrograph GIANO during two transits of the planet.}
{In contrast to absorption in the Earth's atmosphere (telluric absorption), the planet transmission spectrum shifts in radial velocity during transit due to the changing orbital motion of the planet. This allows us to remove the telluric spectrum while preserving the signal of the exoplanet. The latter is then extracted by cross-correlating the residual spectra with template models of the planet atmosphere computed through line-by-line radiative transfer calculations, and containing molecular absorption lines from water and methane.}
{By combining the signal of many thousands of planet molecular lines, we confirm the presence of water vapour in the atmosphere of \pname\ at the 5.5-$\sigma$ level. This signal was measured only in the first of the two observing nights. By injecting and retrieving artificial signals, we show that the non-detection on the second night is likely due to an inferior quality of the data. The measured strength of the planet transmission spectrum is fully consistent with past CRIRES observations at the VLT, excluding a strong variability in the depth of molecular absorption lines.}
{}


\maketitle

\section{Introduction}

In the past few years, spectroscopy at resolving powers above 20,000 has become a valuable tool to unravel the composition, structure, and dynamics of exoplanet atmospheres. The key aspect of this technique is its ability to resolve molecular bands into the individual lines. This enables line-by-line comparison of model spectra to observations, typically obtained by cross-correlation. Hundreds or thousands of molecular lines are co-added into a single cross-correlation function (CCF), enhancing their faint signal by a factor of approximately  $\sqrt{N_\mathrm{lines}}$.
High spectral resolution also allows us to resolve the change in the orbital radial velocity of the planet. For exoplanets orbiting their parent star in just a few days, these changes are in the order of tens of \kms\ across a few hours. Thanks to its unique and highly variable Doppler signature, the spectrum of the planet can be disentangled from the main contaminants (i.e. the spectrum of the Earth's atmosphere and the stellar spectrum), which are instead stationary or quasi-stationary. Thus, detections at high spectral resolution are particularly robust against contamination from spurious sources. Furthermore, they are self-calibrated, eliminating the necessity of a reference star of similar brightness and spectral type in the same field of view.

After more than a decade of inconclusive attempts at using high-resolution spectroscopy for investigating exoplanet atmospheres \citep{col99, cha99, wie01, lei03, luc09, bar10, rod10}, \citet{sne10} devised a successful observational strategy and detected CO absorption lines in the transmission spectrum of exoplanet HD~209458~b via near-infrared (NIR) spectroscopy with the Cryogenic InfraRed Echelle Spectrograph \citep[CRIRES,][]{crires} at the ESO Very Large Telescope (VLT). Two years later, \citet{bro12} and \citet{rod12} demonstrated that the same measurements could be extended to non-transiting planets and used to estimate their orbital inclination and hence true mass. While CRIRES began to routinely detect CO and \water\ in the atmospheres of transiting \citep{dek13, bir13} and non-transiting \citep{bro13, bro14, bir17} planets, \citet{loc14} showed that similar measurements were possible even at a lower resolving power of 25\,000, obtained with NIRSPEC at Keck. They designed an alternative analysis where the planet's Doppler shift no longer needs to change during an observing night \citep{pis16, pis17}, potentially enabling the study of longer-period bodies.
Besides detecting molecular species, determining their relative abundances, and constraining the overall thermal structure of the atmosphere, high-resolution spectroscopy is also capable of measuring planet rotation and even global winds \citep{sch16, bro16} thanks to the broadening and asymmetry in the cross-correlation function.

So far the major downside of the method has been the enormous request in terms of signal-to-noise. Only the biggest telescopes on the ground, namely VLT and Keck, have succeeded in providing an adequate collective area for securing solid detections in the near-infrared, and only for the brightest known exoplanet systems (stellar K magnitudes below 7.5). This has limited the sample to a handful of hot Jupiters orbiting very bright stars with spectral types ranging from late F to early K.
This limitation is, however, tied to the technology of near-infrared spectrographs that were available in the past, and can be overcome by the recent development of instruments with bigger throughput and/or larger spectral range. The high resolution spectrograph GIANO \citep{giano1, giano2}, mounted at the Nasmyth-A focus of the 3.6-m Telescopio Nazionale Galileo (TNG), belongs in the latter category. With its nearly contiguous coverage of the entire Y, J, H, and K bands, it provides a 21-fold increase in spectral range compared to CRIRES at the VLT. 

In this work we have successfully tested the capabilities of GIANO for characterising exoplanet atmospheres. We analysed spectra of \pname\ \citep{bou05} taken in and around transit and extract the signal of the transmission spectrum of the planet. \pname\ is considered one of the best-studied exoplanets to date, due to the brightness of its parent star (K = 5.54). Therefore it serves as the ideal testbed for the scalability of high-resolution spectroscopy to smaller telescope facilities. 

\section{The GIANO spectra and their calibration}

Two transits of \pname\ were observed on July 11 and July 30, 2015, with GIANO at the TNG. The data consist in a sequence of nodded observations following an ABAB pattern, each exposed for 60 s. At each nodding position, a fibre of 1$\arcsec$ in diameter (on-sky) feeds the spectrograph. A second fibre, located at the other nodding position (3$\arcsec$ away) is instead looking at the sky, providing an accurate reference for subtracting the thermal background and telluric emission lines. Each time the telescope nods, the fibres swap, i.e. the object fibre becomes the sky fibre and vice-versa.

The fibres were re-imaged onto a $2\times$ slicer, resulting in two spectral tracks per nodding position, or four tracks per order. The entire NIR spectrum (0.95--2.45~\micron) was cross-dispersed at a resolving power of $R=50\,000$ and orders 32 to 80 are imaged on a 2k $\times$ 2k HAWAII-2 detector.
Contiguous wavelength coverage is ensured for orders higher than 45 (bluer than $\sim1.7$~\micron), whereas for lower orders (redder wavelengths) the coverage is incomplete, down to 75\% for order 32. In this work, we have renumbered the orders so that order 0 is the reddest and order 46 the bluest. We skipped the extraction of the two bluest orders in the cross-dispersed spectrum for lack of photons.

\begin{figure}[t]
\centering
\includegraphics[width=8.5cm]{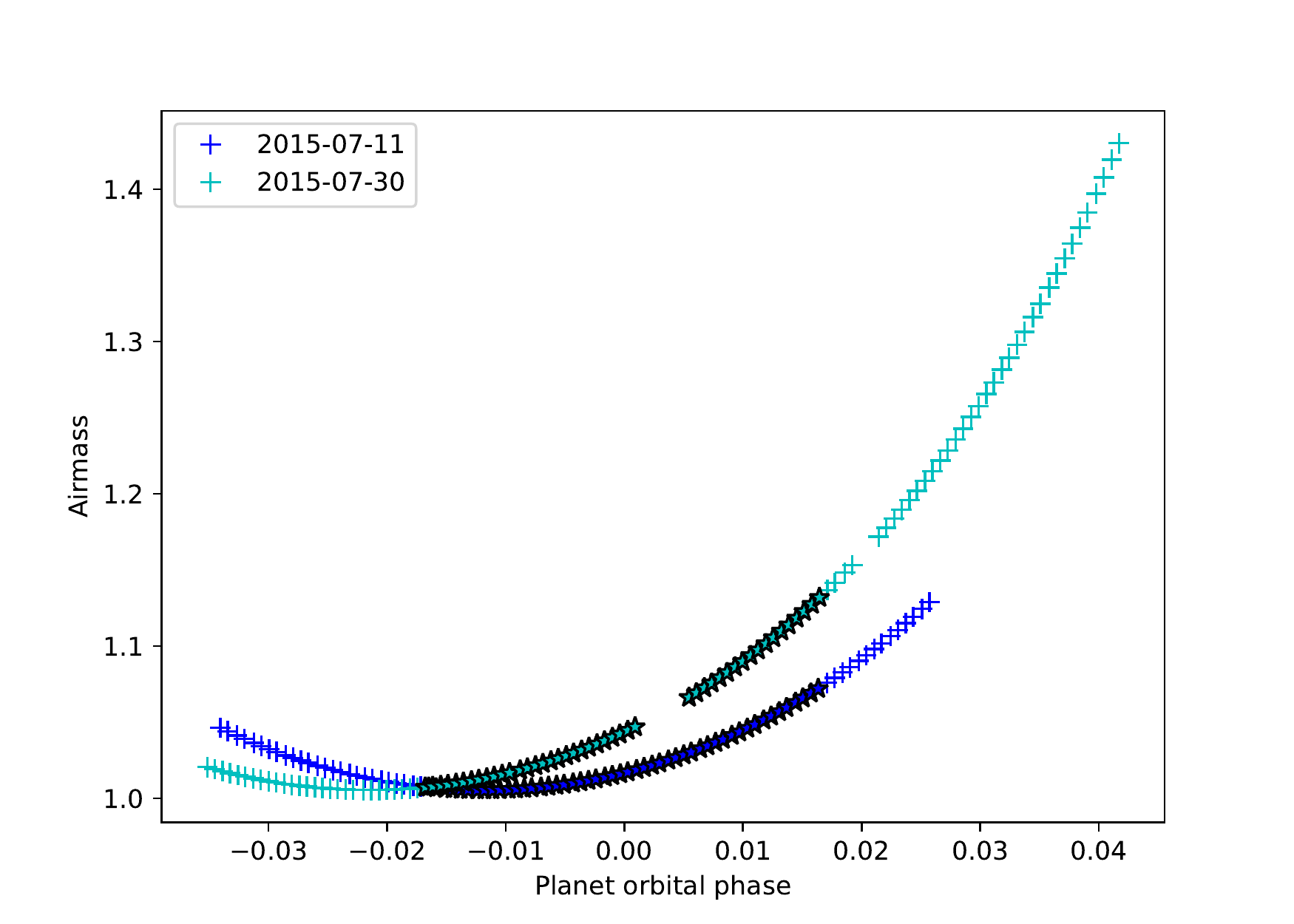}
\caption{Orbital phase of exoplanet \pname\ and corresponding airmass during the two nights of observations analysed in this work. Plus and star symbols denote out-of-transit and in-transit data, respectively.}
\label{fig:ph-air}
\end{figure}

Both observations consist of spectra taken before, during, and after the transit of the planet (Figure~\ref{fig:ph-air}). We collected a total of 90 and 110 spectra in the night of July 11 and 30, respectively. We measured a signal-to-noise ratio (S/N) of 50-60 per spectrum per pixel averaged across the entire spectral range. Peak values of S/N = 70-80 were reached in the H band and correspond to particularly transparent regions of the Earth's atmosphere.

\subsection{Extracting the one-dimensional spectra}

Each science spectrum is processed with our implementation of the IDL algorithm {\tt fixpix}\footnote{\url{https://www2.keck.hawaii.edu/inst/nirspec/redspec_man.pdf}} with 10 iterations to correct for bad pixels. Difference images are created by subtracting each B-image (i.e. the spectrum taken in nodding position B) from the A-image, after dividing each image through the master flat field. This subtraction removes the thermal background and sky emission lines.

Differently from common practice, we did not merge the A and B spectra into a combined AB spectrum. Conversely, we extracted the spectra separately at the A and B positions to increase the time resolution of our observations. This choice also overcomes the necessity of measuring the shift between the dispersion solution in A and B spectra at this stage. This step would be redundant since we re-align the spectra to the telluric reference frame in Section~\ref{analysis:alignment}.

We used the master flat-field to identify the apertures corresponding to each order, each nodding position, and each sliced spectral trace. On the (A--B) flat-fielded science images, we refined the position of the apertures by fitting the spectral trails with four Gaussian profiles per order simultaneously (two sliced images and two nodding positions) at 32 equally-spaced positions along the dispersion direction. These sampled values were then fitted with a second-order polynomial to the full 2048 pixels of the dispersion direction and for each pixel the spectral traces are rectified in the spatial direction via spline interpolation. Lastly, the extraction of the one-dimensional spectra is performed by optimal extraction \citep{hor86} on the rectified images. B spectra need to be multiplied by $-1$ to correct for the initial subtraction of the B images. The count levels of each spectrum are also rescaled to that obtained with a rectangular extraction. This is useful for S/N estimates, see Section~\ref{sec:discussion}.

\begin{figure*}[ht]
\centering
\includegraphics[width=18cm]{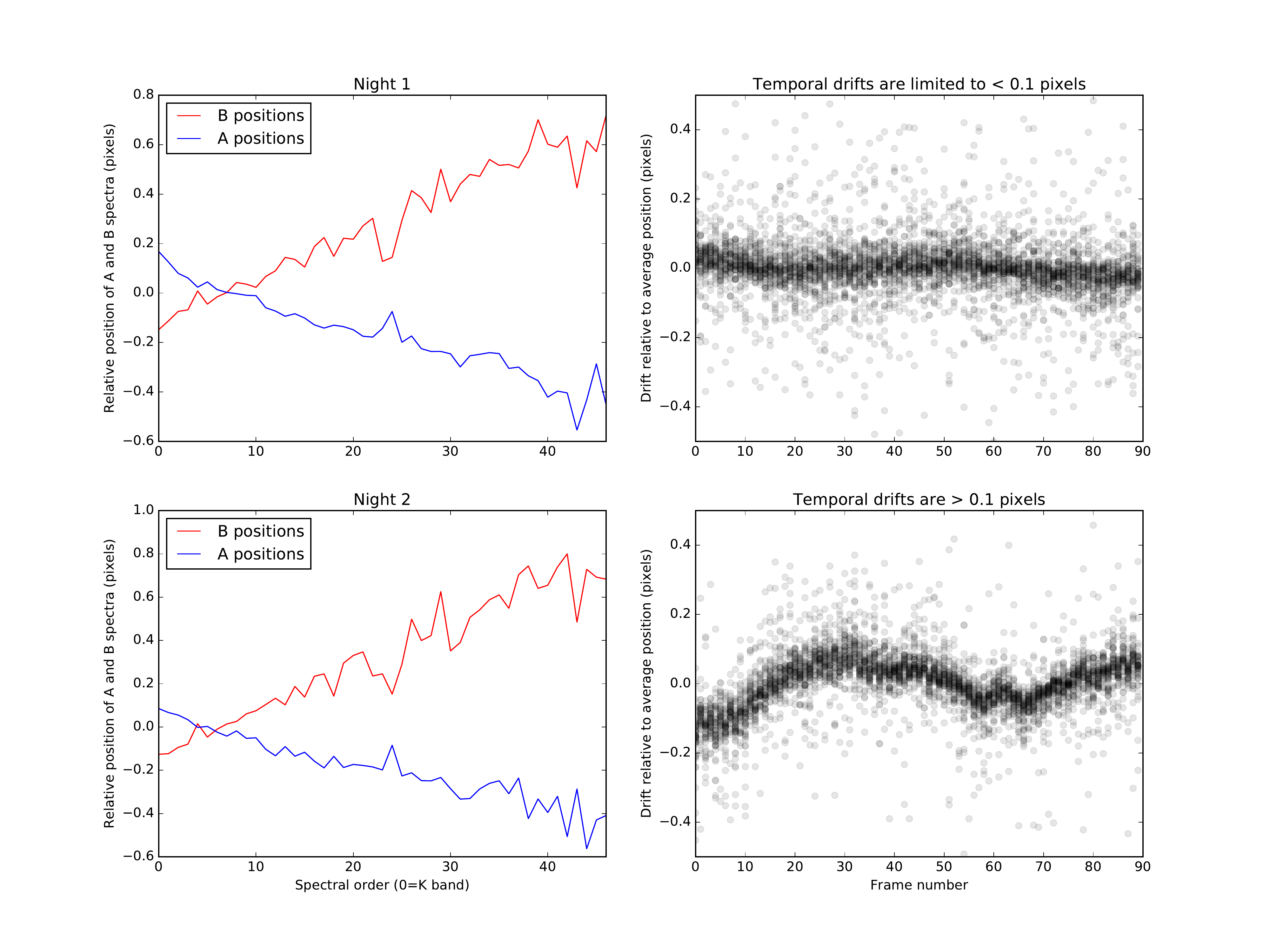}
\caption{Stability of the GIANO spectrograph. The dispersion solution of the spectra taken at A and B position differs as a function of the spectral order considered (left panel, averaged over the observing night). In addition, we measure temporal variations in the global dispersion solution (right panel). Their amplitude differs between the first and the second night of observations (top-right and bottom-right panels, respectively). These measurements suggest that although A and B spectra move coherently (i.e. their relative shift stays the same), their absolute position is variable on a timescale of minutes. Furthermore, the overall stability of GIANO differs between observing nights.}
\label{fig:shifts}
\end{figure*}

\subsection{Aligning the spectra to a common reference frame}\label{analysis:alignment}
The high spectral resolution of GIANO allows us to resolve the orbital motion of the planet \pname\ during transit. Its radial velocity indeed changes by approximately 32~\kms\ from ingress to egress. In contrast, the absorption spectrum of the Earth's atmosphere (telluric spectrum) is stationary in wavelength. We intend to exploit this intrinsic difference in the Doppler signature of the planetary and telluric signals to disentangle the former from the latter. As part of this strategy, we align all the observed spectra to the telluric reference frame.

We measured the position of a set of telluric lines in every order and every spectrum in the sequence. The left-hand panels in Figure~\ref{fig:shifts} show that there is a clear trend in wavelength between the dispersion solution for the A positions (in red) and that of the B positions (in blue). This trend seems consistent between the two observing nights (top and bottom row). The right-hand panels show the temporal variations in the wavelength solution once the overall trend is removed. Not only does the wavelength recorded on a certain pixel of the detector vary as a function of time, but also the amplitude of this variation differs between the two observing nights. Whereas on July 11 the dispersion solution of GIANO is stable to 1/20 of a pixel (1 pixel equals to 2.7 \kms), on July 30 it varies by at least 0.2 pixels. We further discuss the stability of the night of July 30 in Section~\ref{sec:2nd_night}. For both nights, we proceeded to re-align each spectrum in the temporal sequence by spline interpolating it based on the measured shifts. 

\subsection{Calibrating the spectra in wavelength}\label{analysis:wcal}
Once the spectra are all in the telluric reference frame, we calibrated their wavelengths. We note that the set of arc frames taken after the observing night are not suitable for calibration at the level of precision required by our analysis. Since we showed in Section~\ref{analysis:alignment} the instability of the dispersion solution in time, any calibrations non-simultaneous to the science observations could result in a biased wavelength solution. 
In the NIR telluric absorption lines provide instead an excellent simultaneous calibration source, except for some orders of the $J$ band where the Earth's atmosphere is particularly transparent. However, the star \sname\ is a K1-2V dwarf and it shows significant absorption lines at near-infrared wavelengths. These can also be used for calibration. 

\begin{figure*}[ht]
\centering
\includegraphics[width=18cm]{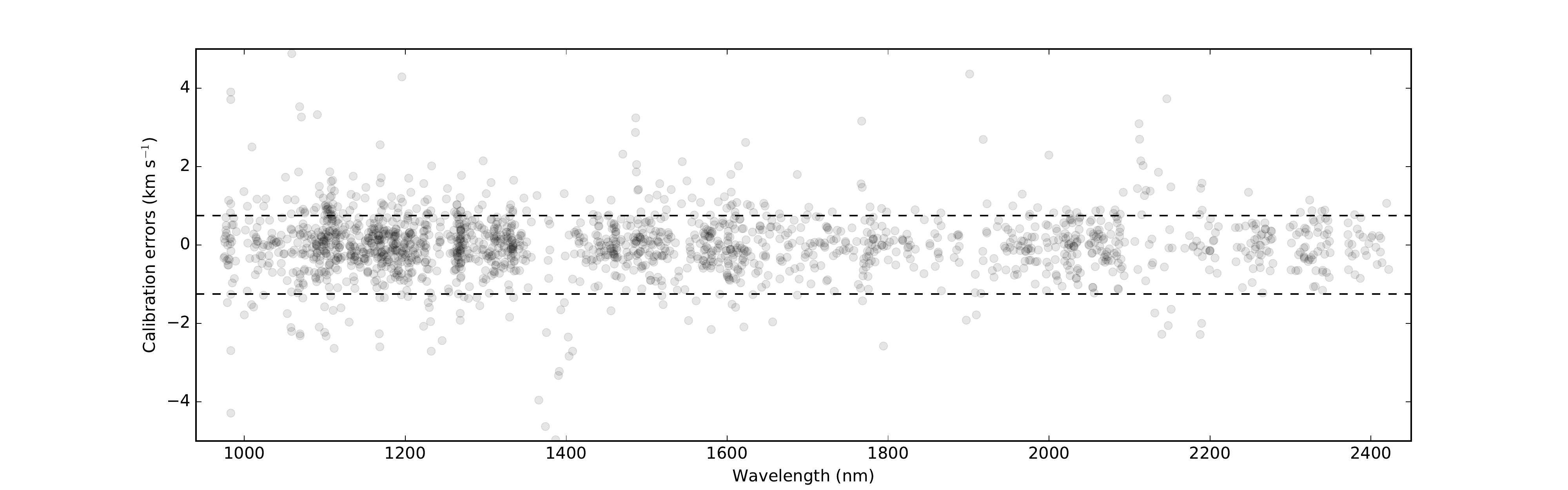}
\caption{Residuals of the wavelength calibration (in \kms) versus wavelength for the full spectral coverage of GIANO. The dashed lines denote the $\pm1\sigma$ in the distribution of the residuals. For most of the orders, a sub-km/s per line wavelength solution is achievable.}
\label{fig:wcal}
\end{figure*}

We built a template for wavelength calibration by multiplying a Phoenix stellar spectrum \citep{all11} matching the properties of \sname\ by the atmospheric transmission spectrum generated via the ESO Sky Model Calculator\footnote{\url{https://www.eso.org/observing/etc/bin/gen/form?INS.MODE=swspectr+INS.NAME=SKYCALC}}. The stellar spectrum is Doppler-shifted based on the barycentric and systemic velocities at the time of the observations. 
For each order, we visually paired absorption lines in the template and in the averaged observed spectrum. We then computed the wavelengths of the line centroids in the template spectrum, and the pixel value of the corresponding lines in the averaged spectrum. The centroids were determined by computing a super-sampled version of each line via spline interpolation. Super-sampling is achieved via spline interpolation at the 1/20 of a pixel. We recorded the fractional pixel at which the flux in the super-sampled observed line reaches a minimum, and the corresponding wavelength in the super-sampled template spectrum. We exclude from the analysis lines that appear saturated, close to saturation, or blended. The (pixel, wavelength) relation obtained from the above is fitted with a fourth-order polynomial, and the fit is assumed as the wavelength solution. For most of the orders we achieve a residual scatter per line well below 1 km s$^{-1}$. Figure~\ref{fig:wcal} shows the residuals as a function of wavelength, and its $+1\sigma$ (0.75 km s$^{-1}$) and $-1\sigma$ (1.25 km s$^{-1}$) dispersion on the global sample.
Some of the orders corresponding to spectral windows where the Earth's atmosphere is particularly opaque do not have enough flux or unsaturated spectral lines to calibrate in wavelength. Furthermore, for a few orders in the J band, we notice an evident mismatch between modelled and observed spectral lines, which also prevents us from obtaining a reliable calibration. Due to these challenges, we exclude six orders from the night of July 11 (orders 8, 9, 10, 23, 24, 25, corresponding to the ranges 983.06-983.30 nm, 1352.7-1408.7 nm, and 1804.8-1937.8 nm) and six orders from the night of July 30 (orders 8, 23, 24, 43, 45, 46, corresponding to 970.8-983.4 nm, 1009.7-1032.3 nm, 1352.8-1408.7 nm, and 1895.1-1937.8 nm). 

\section{Extracting the planet signal}
At this stage of the analysis the planet signal is outshone by orders of magnitude by the stellar and telluric spectra. However, the planet's orbital velocity has a non-zero radial component during transit, which amounts to $-16$~\kms\ in ingress and $+16$~\kms\ in egress. Consequently, while telluric and stellar lines are stationary or quasi-stationary (the stellar barycentric velocity changes by 0.2~\kms\ during transit) in wavelength, the planet spectrum experiences a detectable change in Doppler shift during the 110 minutes of transit. The analysis described in this section aims to make use of this peculiar Doppler signature to remove the telluric and stellar spectra while preserving the planet signal as much as possible. 

\subsection{Removing the Earth's absorption spectrum}\label{sec:tell_corr}
We have treated each order separately, as a bidimensional matrix in which wavelength ($\lambda$) is on the $x$-axis and time ($t$) on the $y$-axis. Due to the possible differences in the properties of the detector (quantum efficiency and bad pixels) between the two nodding positions, A and B spectra (i.e. odd and even spectra in the temporal sequence) are also treated separately. Lastly, we report a clear discontinuity between the right and left quadrant of the detector. Hence, we also separated $x$-pixels 0--1023 from 1024--2047. We therefore process $41\times2\times2$ = 164 matrices of spectra for each observing night. We note here that we already discarded the six orders with inaccurate wavelength calibration (Section~\ref{analysis:wcal}).

The analysis begins by taking the natural logarithm of each matrix. This choice reflects our expectation that at first order the depth of telluric lines depends on the exponential of the geometric airmass ($a(t)$, see below). Before taking the logarithm, all the pixels below a linear count level of 0.1 are masked to avoid divergent or infinite values. Residual bad columns of data (i.e. certain wavelengths falling on damaged portions of the detector), previously identified by visual inspection, are also masked. 

With the data in logarithmic space, the median of the 300 brightest pixels is subtracted from each observed spectrum to remove variations in throughput due to pointing, seeing, and sky transparency. A temporal average of each matrix is then taken to construct the mean observed spectrum for the night. A robust linear fit between each observed spectrum and the mean spectrum is computed and subtracted out, removing most of the spectral features. However, an additional correction is required to model the change with time of the depth of telluric lines. For each $\lambda_i$ in the data, this is done by subtracting the fit of the observed flux $\log{[F(\lambda_i,t)]}$ with the geometric airmass $a(t)$. If all the species in the Earth's atmosphere were vertically mixed and had constant abundances, a linear fit with airmass would suffice. However, the water vapour content is likely to change during the night, and in addition water is confined to the Earth's troposphere. Consequently, the amount of water vapour along the line of sight will change in a non-trivial fashion. In these data, we fitted the functional dependence by airmass as
\begin{equation}
\log{[F(\lambda_i,t)]} = c_{i0}+ c_{i1}a(t) + c_{i2}a^2(t),
\end{equation} 
which we find far superior to a linear fit in modelling the depth of telluric lines. The fit is computed and subtracted out for each of the spectral channels in the data (each of the columns in the ($\lambda, t$) matrix). Although we still detected residual telluric signal in our data (see Section~\ref{sec:ccorr} below), we verify that it is not possible to improve the correction by just increasing the degree of the polynomial in $a(t)$. We suspect that this happens because even on a photometric night the water vapour content changes irregularly on timescales of minutes, and these variations cannot be easily modelled with a smooth function of time.

With telluric lines removed as above, the previous mask for bad columns and/or low-count pixels is applied again, setting the corresponding pixels to zero. An additional sigma-clipping is applied with threshold of eight times the standard deviation of each matrix to ensure that no residual strong outlier is present. Lastly, the exponential of the data is taken to restore linear flux values.

Since we removed only the time-averaged quantities from the data, or we modelled each spectral channel as function of time separately, we preserved most of the planet signal in the process. This is due to the fact that absorption lines in the planet spectrum will shift wavelength during the night, and therefore will fall on a certain spectral channel only for a limited amount of time. 

\subsection{Enhancing the planet signal via cross-correlation}\label{sec:ccorr}
Even if preserved, the planet spectrum is still below the noise threshold at this stage. The deepest absorption lines in the planet transmission spectrum barely reach a depth of 1\% compared to the stellar continuum, while our typical S/N is 50-60 per spectrum per spectral channel. However, there are thousands of molecular lines in the spectral range covered by GIANO. Their signal can be co-added by cross-correlating the residual data after Section~\ref{sec:tell_corr} with template spectra for the planet atmosphere. These are computed via plane-parallel, line-by-line radiative transfer calculations following the prescriptions of \citet{bro16} for what concerns $T$-$p$ profile and H$_2$--H$_2$ collision-induced absorption. In this work CO is not included as trace gas. Given that this molecule mainly absorbs at selected wavelength intervals (longer than 2290 nm, and much more weakly in the range 1560-1600 nm), that GIANO data is affected by strong modal noise (i.e. noise related to fibre modes) in the $K$ band, and that the stellar Rossiter-McLaughlin effect would likely dominate the CO signal without an appropriate correction of the stellar spectrum \citep{bro16}, we opt to explore \water\ and \methane\ as trace gases in this analysis, with opacities taken from HITEMP2010 \citep{rot10} and HITRAN2012 \citep{rot12} respectively. Not only are these two species strong tracers of chemistry in hot-Jupiter atmospheres \citep{mos11, mad12, mos13}, but they also show a dense forest of spectral lines across most of the NIR. Hence, they are well suited to exploit the large spectral range of GIANO.

In our calculations we first set the volume mixing ratio (VMR) of methane to $10^{-12}$ and explore VMRs of  $10^{-5}$, $10^{-4}$, and $10^{-3}$ for water. We then set VMR(\water) = $8\times10^{-4}$, i.e. the chemical equilibrium value for $T=1500$~K, solar C/O, and solar metallicity \citep{mad12}, and add \methane\ at VMRs between $10^{-7}$ and $10^{-4}$, in increments of one order of magnitude. We kept the $T$-$p$ profile in the lower part of the atmosphere ($p > 0.1$ bar) fixed and constrained to results from past theoretical and observational studies \citep{mad09}. However, we have explored two different temperatures (500 K and 1500 K) for the upper part of the atmosphere. Although these would produce completely different emission spectra (i.e. absorption and emission lines, respectively), they have a much smaller impact on the transmission spectrum. The main effect is on the overall depth of absorption lines, with the hotter upper atmosphere producing models with up to 4$\times$ the contrast ratio of the colder models. The influence on the relative strength of absorption lines is instead minimal. In conclusion, for this study we test a set of 14 models (seven relative abundances and two $T$-$p$ profiles).

Each model has the average transit depth subtracted before cross-correlation. In this way both the models and the data have their continuum set to zero. The cross correlation is computed on a fixed grid of radial velocity lags between $-225$ and $+225$~\kms, in steps of 2.7~\kms, which approximately matches the pixel scale for most of the GIANO orders. This results in a lag vector of 167 radial velocity values.
For each spectrum, each order, and each radial-velocity lag, the Doppler-shifted model spectrum is computed by spline interpolation and cross-correlated with the data. 
The output of the cross-correlation is a set of 47 matrices (one per order), each with dimensions ($167 \times n_\mathrm{spec}$), where $n_\mathrm{spec}$ is the number of spectra on each night (90 and 110 on July 11 and 30, respectively). 
\begin{figure}[th]
\centering
\includegraphics[width=8.5cm]{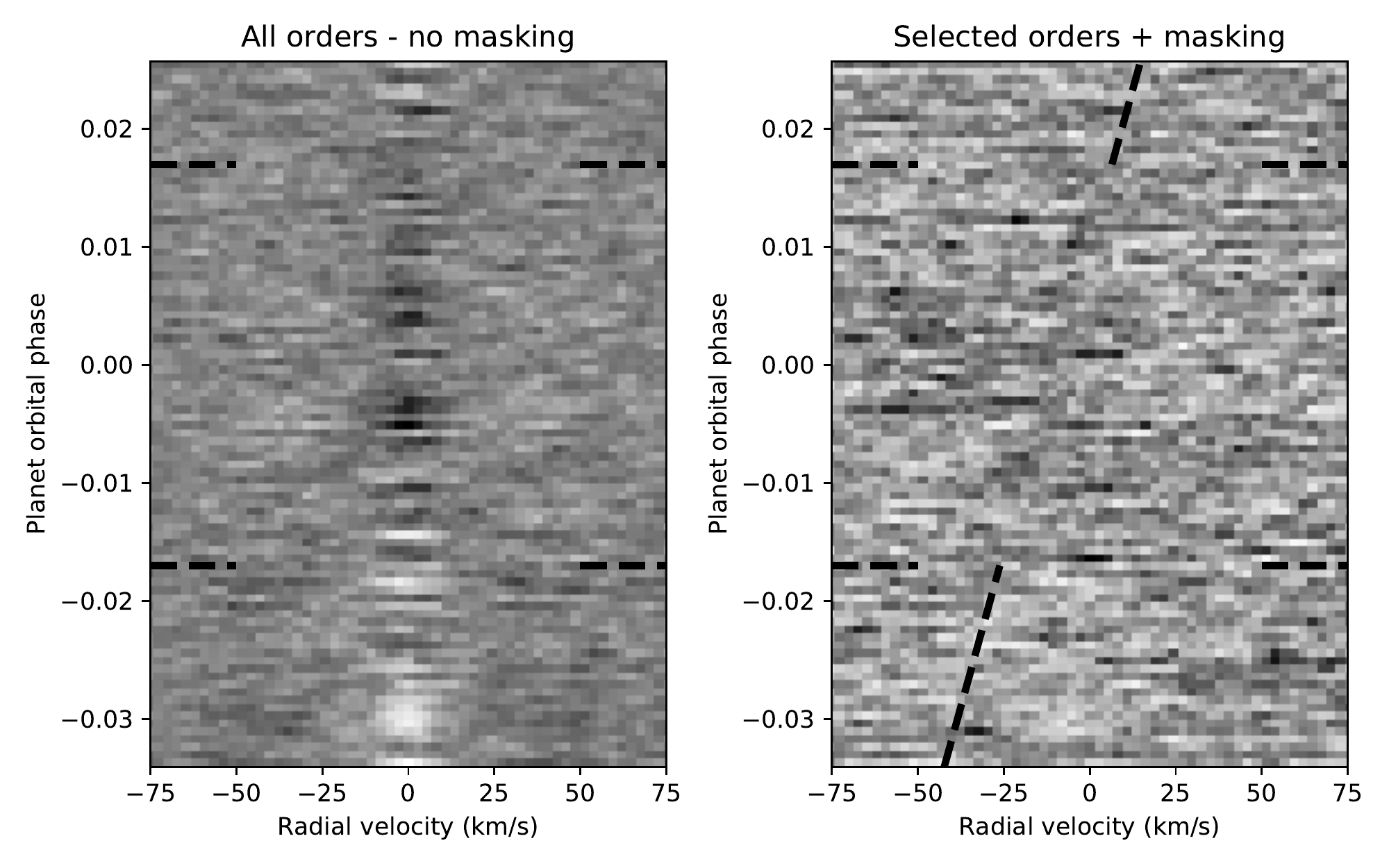}
\caption{Cross-correlation between the sequence of spectra of \sname\ observed with GIANO on July 11, 2015, and a model spectrum for the planet \pname\ containing \water\ at VMR = $10^{-4}$, shown as function of radial velocity and planet orbital phase. Ingress and egress of \pname\ are marked with horizontal dashed lines. In the left panel, we co-added all the GIANO orders after removing telluric lines similarly to past CRIRES observations \citep{bro14, bro16}. In the right panel, we applied an additional mask at the position of telluric \water\ lines. The latter step removes the majority of time-correlated residual telluric noise, leaving a faint trail of positive correlation along the expected planet radial velocity (marked by slanted dashed lines). Details can be found in Section~\ref{sec:tell_corr}.}
\label{fig:trails}
\end{figure}

Even at this stage the planet signal is not expected to be detectable yet. We therefore proceeded to co-add the 47 cross-correlation matrices with equal weights, even though the signal in each order will depend on i) the transparency of the Earth's atmosphere, ii) the density and depth of absorption lines in the spectrum of \pname, and iii) the position of exoplanetary lines compared to strong telluric lines. Factors ii) and iii) could be computed from our set of model spectra used for cross-correlation, but this would make the weighting model-dependent. In order to avoid this bias, we therefore assumed equal weighting. 

The remainder of this Section describes the analysis tuned specifically for the night of July 11, 2015. This is due to the fact that on the night of July 30, 2015, no signal is detected (see Section~\ref{sec:results} and more specifically Section~\ref{sec:2nd_night}). However, the qualitative behaviour of the cross-correlation signal and the consequent strategy, discussed below, apply to both nights. As illustrated in the left panel of Figure~\ref{fig:trails}, the co-added cross-correlation signal shows a prominent feature spanning the entire sequence and smoothly shifting in time from anti-correlation (lighter colours in the plot) to correlation (darker colours). We identified this feature with spurious residual telluric absorption because it is centred at zero radial-velocity lag in the telluric frame, it appears regardless of the planet orbital phase (hence it is not correlated with the transit of the planet), and it is enhanced when cross-correlating the data with a model for the Earth's transmission spectrum (not shown here), containing the cold \water\ spectrum rather than the more complex and rich hot \water\ spectrum of the planet \pname. 

In past CRIRES data where we de-trended telluric lines by airmass \citep{bro12, bro13, bro14}, residuals were clearly visible in the processed spectra prior to cross-correlation (i.e. at the end of the analysis in Section~\ref{sec:tell_corr}). Therefore, the strongest measured residuals at the position of telluric lines could be used to de-trend the rest of the spectral channels, generally achieving a near photon noise correction. In these GIANO data, however, no residual is clearly visible prior to cross-correlation. We therefore took a more straightforward approach where we mask the data prior to cross-correlation by assigning zero value to those spectral channels corresponding to telluric lines deeper than 5-30\%, with the actual threshold varying based on the order but not significantly influencing the outcome. We repeated the cross correlation, and visually inspect the matrices for each order. For 24 of the 41 orders with good wavelength calibration, any residual telluric signal is suppressed below the cross-correlation noise. For the remaining 17 orders, we still identified some cross-correlation noise by visual inspection, although in broader patterns spanning tens of \kms\ in radial velocity. These noise structures do not resemble the signature of residual telluric lines. We therefore also discarded these additional orders from the analysis, ending up utilising approximately 50\% of the available spectral range of GIANO (orders 2, 7, 12-17, 19, 27-30, 33-35, 38-44, and 46). We note that some of the discarded orders are affected by either strong telluric absorption (orders 8-10, 22-25) or modal noise (orders 0-6). They are both likely causes of this extra cross-correlation noise.

With the 24 good orders selected as above, we achieved a co-added matrix of CCFs relatively free from telluric contamination (Figure~\ref{fig:trails}, right panel). A slanted trail of darker cross-correlation values (higher correlation in the adopted colour scheme) is barely visible at the expected planet radial velocity (along the slanted guidelines) and only in-transit. This is the CCF of the planet transmission spectrum, which we then proceeded to co-add in the rest-frame of \pname\ to obtain the total signal from the planet.

\begin{figure}[ht]
\centering
\includegraphics[width=8.5cm]{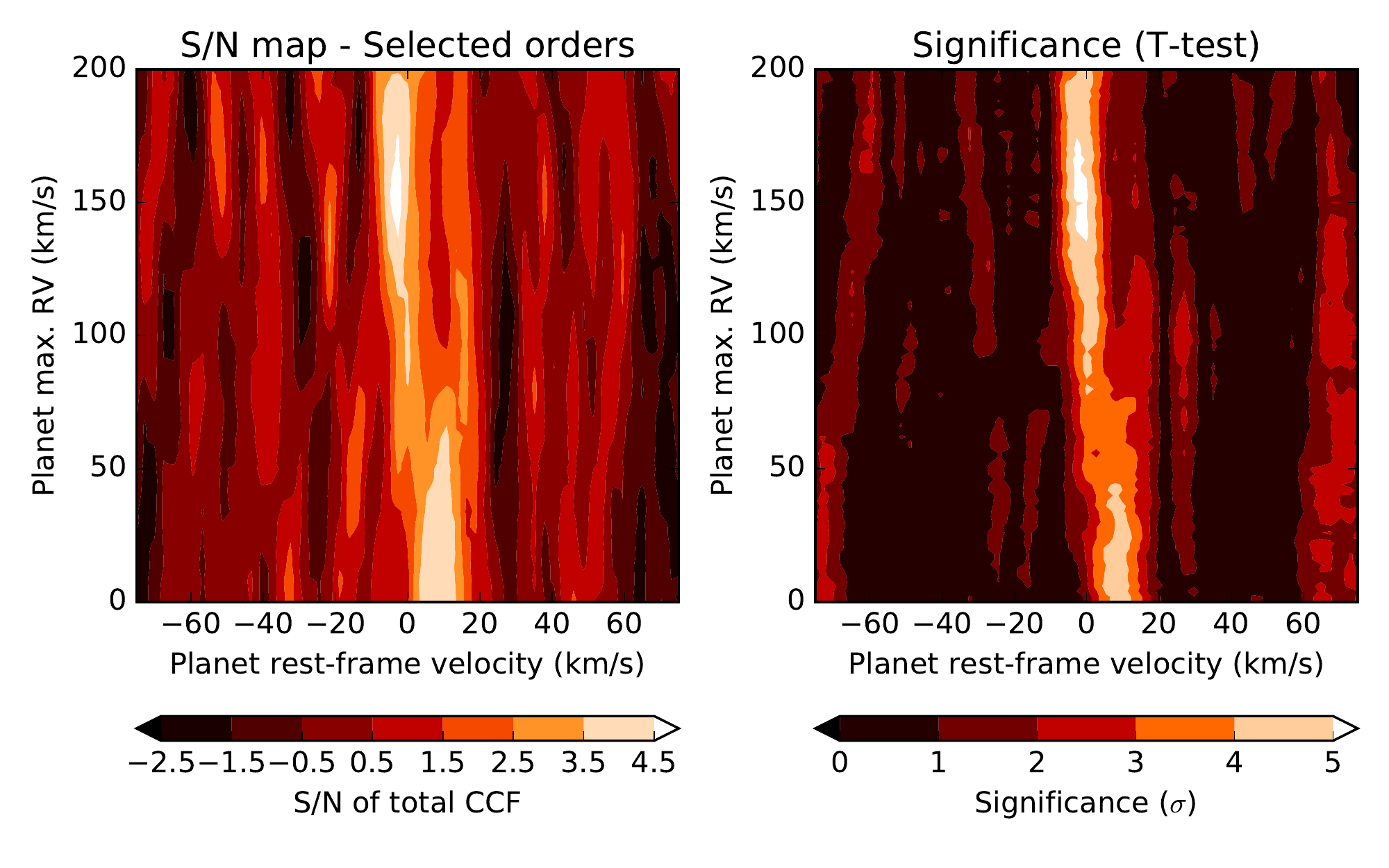}
\includegraphics[width=8.5cm]{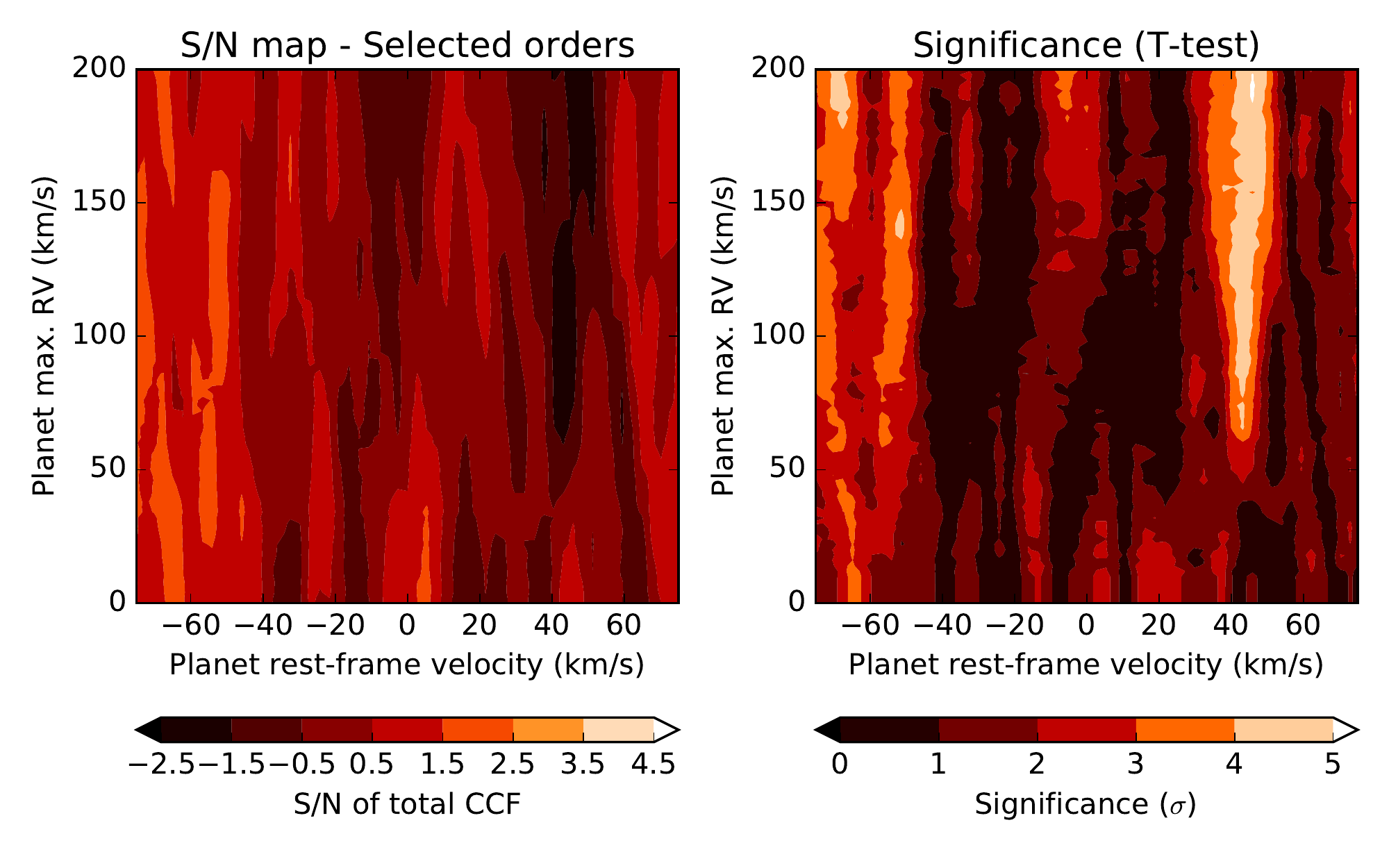}
\caption{Detection of \water\ in the transmission spectrum of \pname\ observed on July 11, 2015 (top panels) and the non detection on the night of July 30, 2015 (bottom panels). The total cross correlation signal is shown as function of rest-frame velocity $v_\mathrm{rest}$ and planet radial velocity semi-amplitude \kp. The measured S/N and corresponding significances are shown in the left and right panels, respectively. These are computed as explained in Section~\ref{sec:significance}. The cross-correlation shown here is obtained with a model spectrum containing \water\ at VMR = $10^{-4}$ and \methane\ at VMR = 10$^{-12}$.}
\label{fig:diags}
\end{figure}

Co-adding requires computing the planet radial velocity $v_\mathrm{P}$ in the telluric reference system. Assuming a circular orbit, this is given by:
\begin{equation}\label{eq:pl_vel}
v_\mathrm{P}(t, K_\mathrm{P}) = v_\mathrm{bary}(t) + v_\mathrm{sys} + K_\mathrm{P} \sin [2\pi\varphi(t)],
\end{equation}
where $v_\mathrm{bary}$ is the velocity of the barycentre of the solar system with respect to the observer, $v_\mathrm{sys}$ is the systemic velocity of \sname\ \citep[$-2.3$~\kms,][]{tri09}, \kp\ is the planet orbital radial velocity semi-amplitude, and $\varphi$ the planet orbital phase. The latter is zero at the centre of the transit, 0.5 at mid-secondary eclipse, and 0.25-0.75 when the planet is in quadrature. It is computed from the time $t$ of the observations, the time of mid-transit $T_0$ and the planet orbital period $P$ \citep{ago10} as the fractional part of
\begin{equation}
\varphi (t) = \frac{t - T_0}{P}.
\end{equation}
For each value of \kp\ we shifted each cross correlation function at time $t$ via linear interpolation so that it is centred around $v_\mathrm{P}$. We then co-added the shifted cross-correlations in time to obtain the total cross-correlation signal from the planet as a function of rest-frame velocity $v_\mathrm{rest}$ and planet radial velocity semi-amplitude \kp\ (Figure~\ref{fig:diags}, left panels). If the planet transmission spectrum is detected, we measure a signal at $v_\mathrm{rest} = 0.00^{+0.19}_{-0.16}$ \kms\ and \kp\ = $152.5^{+1.3}_{-1.8}$ \kms. The latter being computed from the semi-major axis, orbital period, and orbital inclination in the literature \citep{tri09, ago10}, and by error propagation. Since the uncertainty in $v_\mathrm{rest}$ varies with orbital phase via Eq.~\ref{eq:pl_vel}, we computed it for every value of $\varphi$ and adopted the mean value of the series. This simulates the shifting and co-adding of the CCF illustrated above.

We note that although \kp\ is generally well known for transiting exoplanets, we still explore a full range of values. This allows us to verify that no other spurious signals produce a significant detection near the planet's position. In Figure~\ref{fig:diags} (top panels), for instance, residual uncorrected signal is still present at very low \kp. This is typically due to either stellar residuals (if at zero rest-frame velocity) or telluric residuals (at non-zero rest-frame velocity, as in this case). Exploring a sufficiently large parameter space offers a strong diagnostic power on all sources of noise and we strongly recommend its implementation every time the signal from the planet can be separated in Doppler space from stellar and telluric lines.

\subsection{Determining the significance of the signal}\label{sec:significance}
\begin{figure}[ht]
\centering
\includegraphics[width=8.5cm]{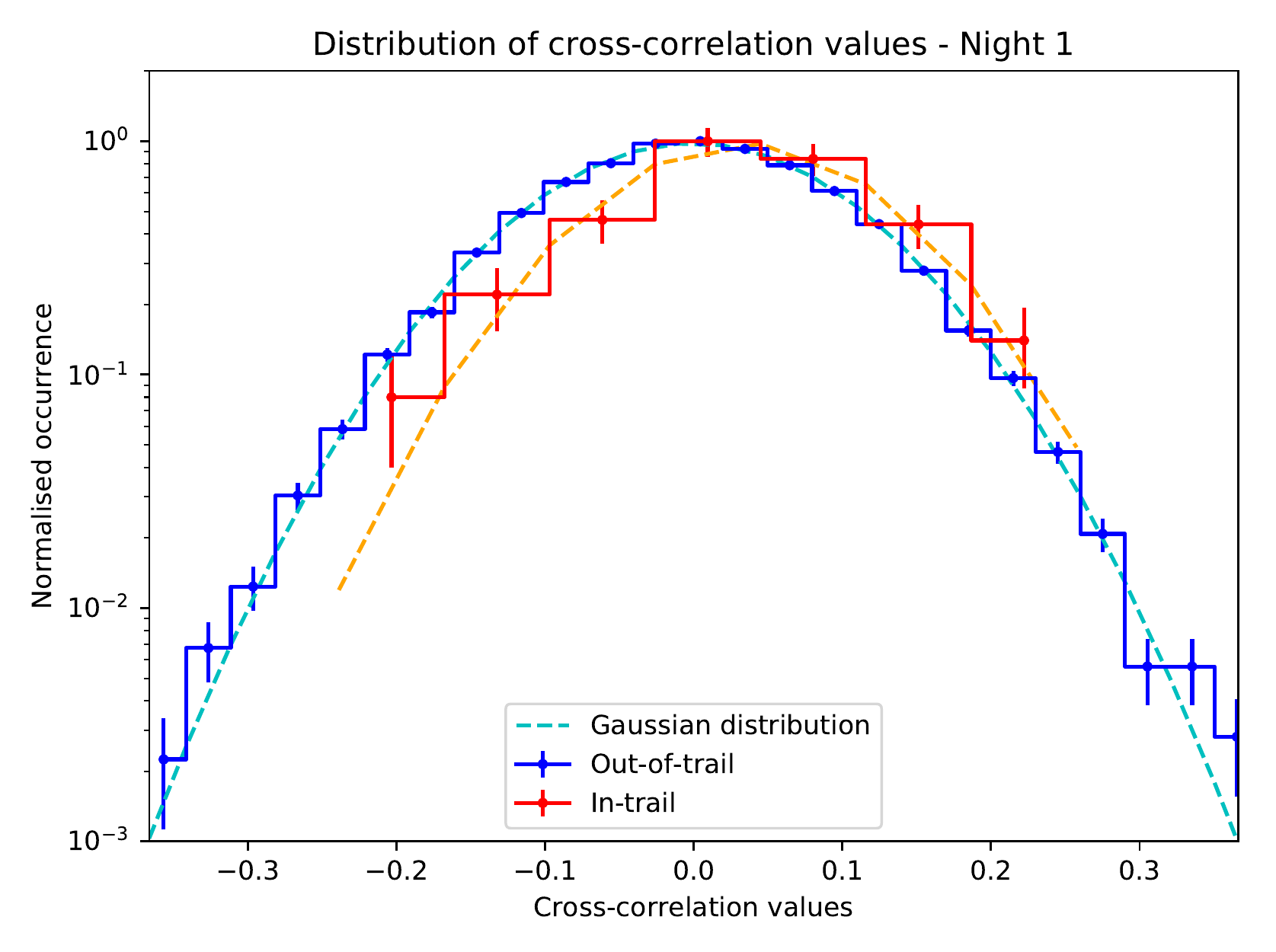}
\includegraphics[width=8.5cm]{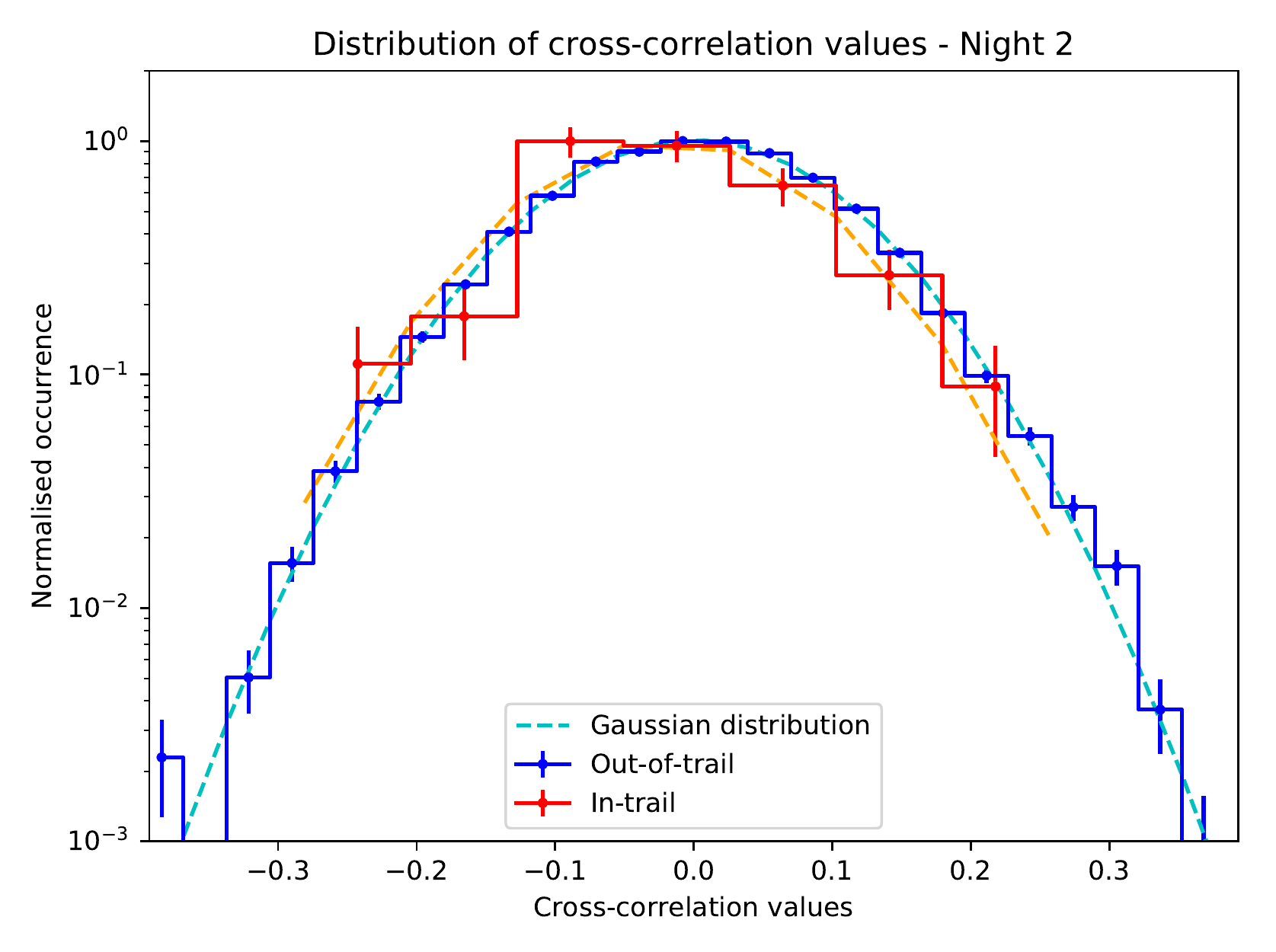}
\caption{Distribution of cross-cross correlation values away from the radial velocity of \pname\ (out-of-trail values, in blue), compared to a Gaussian distribution with the same sample mean and variance (dashed line, light blue). The top panel is for the first night of observations, the bottom panel for the second. In both datasets, the sample distribution shows no deviation down to approximately four times the standard deviation. For comparison, the corresponding distribution of the cross-correlation values around the planet radial velocity (in-trail) is shown in red. In the data from the first night, the two distributions appear shifted as one would expect if the planet transmission spectrum is detected. Conversely, no evident shift is detected in the data from the second night. The statistical significance of these shifts is quantified through a generalised t-test as explained in Section~\ref{sec:significance}.}
\label{fig:cc_distr}
\end{figure}

Following \citet{bro12, bro13, bro14}, we estimated the significance of the detection with two different methods. Firstly, we divided the peak value of the cross-correlation function at each \kp\ by the standard deviation of the noise away from the peak. In this way we computed the S/N of the detection, which is a good proxy for its significance in the absence of significant correlated noise. 
The latter hypothesis is verified by studying the distribution of the cross-correlation values more than 10 \kms\ away from the planet radial velocity (hereafter denoted as `out-of-trail' values). In Figure~\ref{fig:cc_distr}, we compare it to a Gaussian curve with the same mean and standard deviation as the sample. We did not measure any deviations from a Gaussian distribution down to approximately 4$\sigma$, the limit being imposed by the number of available cross-correlation values. In the same Figure, we show in red the distribution of the cross-correlation values within 4 \kms\ from the planet radial velocity (`in-trail' values). It is already evident even from visual inspection that while for the first night (top panel) the in-trail distribution has a higher mean that the out-of-trail distribution (consistent with a higher degree of correlation), in the second night (bottom panel) the two distributions broadly overlap.

We quantified the above statement statistically by performing a generalised t-test on the data. The null hypothesis $H_0$ is that out-of-trail and in-trail values have the same mean. The test is performed for the same range of $v_\mathrm{rest}$ and \kp\ as before. For each value of the two parameters, different sets of cross-correlation values populate the in-trail or out-of-trail sample, and we translated the corresponding t value into a significance at which $H_0$ is rejected (Figure~\ref{fig:diags}, right panels), which is by definition the significance of our detection.

\section{Results}\label{sec:results}

Despite the exclusion of a fraction of the data due to telluric residuals, we clearly detect the absorption of water vapour in the transmission spectrum of \pname\ taken on July 11, 2015. When cross correlating with a model containing \water\ at VMR = $10^{-4}$ and \methane\ at VMR = 10$^{-12}$, we measure a S/N = 4.9 and a significance of $5.5\sigma$ (Figure~\ref{fig:diags}, top panels). Varying the VMR of \water, as well as including non-negligible VMRs of \methane\ in our models result in a marginal decrease in S/N for the entire range of VMRs tested. This suggests that \methane\ is not detected in the planet's transmission spectrum, in line with previous measurements in the $K$ band \citep{bro16}. We do not detect the transmission spectrum of \pname\ on the second available transit (July 30, see Figure~\ref{fig:diags}, bottom panels). In Section~\ref{sec:2nd_night} we further discuss this non-detection.

By averaging the maxima in the CCF and t-test matrices (Fig.~\ref{fig:diags}, top panels), we derived a planet maximum radial velocity of \kp\ = $154^{+36}_{-40}$ \kms. The 1-$\sigma$ uncertainty corresponds to a drop of one in either S/N or significance. The measured \kp\ is fully compatible with the literature value reported in Section~\ref{sec:ccorr} (\kp = $152.5^{+1.3}_{-1.8}$ \kms), and also with the value measured in \citet{bro16}, that is \kp\ = $194^{+19}_{-41}$ \kms. We note that the large uncertainty in \kp\ is due to the relatively small change in planet redial velocity during transit (a few tens of \kms), insufficient to tightly constrain the orbit at this level of S/N. Past dayside observations, particularly those co-adding spectra spanning a wide range orbital phases \citep{bro12, bro14} provided instead much tighter constraints on \kp, with error bars of a few \kms.
We measured a net blue-shift of the planet CCF ($-1.6^{+3.2}_{-2.7}$ \kms). Although very marginal, this is in line with previous measurements by \citet{bro16} and \citet{lou15} regarding the strength of day-to-night side winds at the planet's terminator. We note that high-altitude winds flowing from the dayside to the night-side hemisphere of hot Jupiters are predicted by global circulation models, and expected to produce a radial-velocity anomaly of 1-2 \kms\ in the peak position of the CCF \citep{mil12, sho13, kem14}. This is in line with the measurement in this paper, although the signal detected with GIANO is both significantly broader and at lower significance than the combined CO+\water\ detection of the transmission spectrum of \pname\ made in the $K$ band with CRIRES \citep[7.6$\sigma$, see][]{bro16}. This contributes to the bigger uncertainties in estimating the net blue-shift from the peak position of the co-added planet signal. 

\begin{figure}[ht]
\centering
\includegraphics[width=8.5cm]{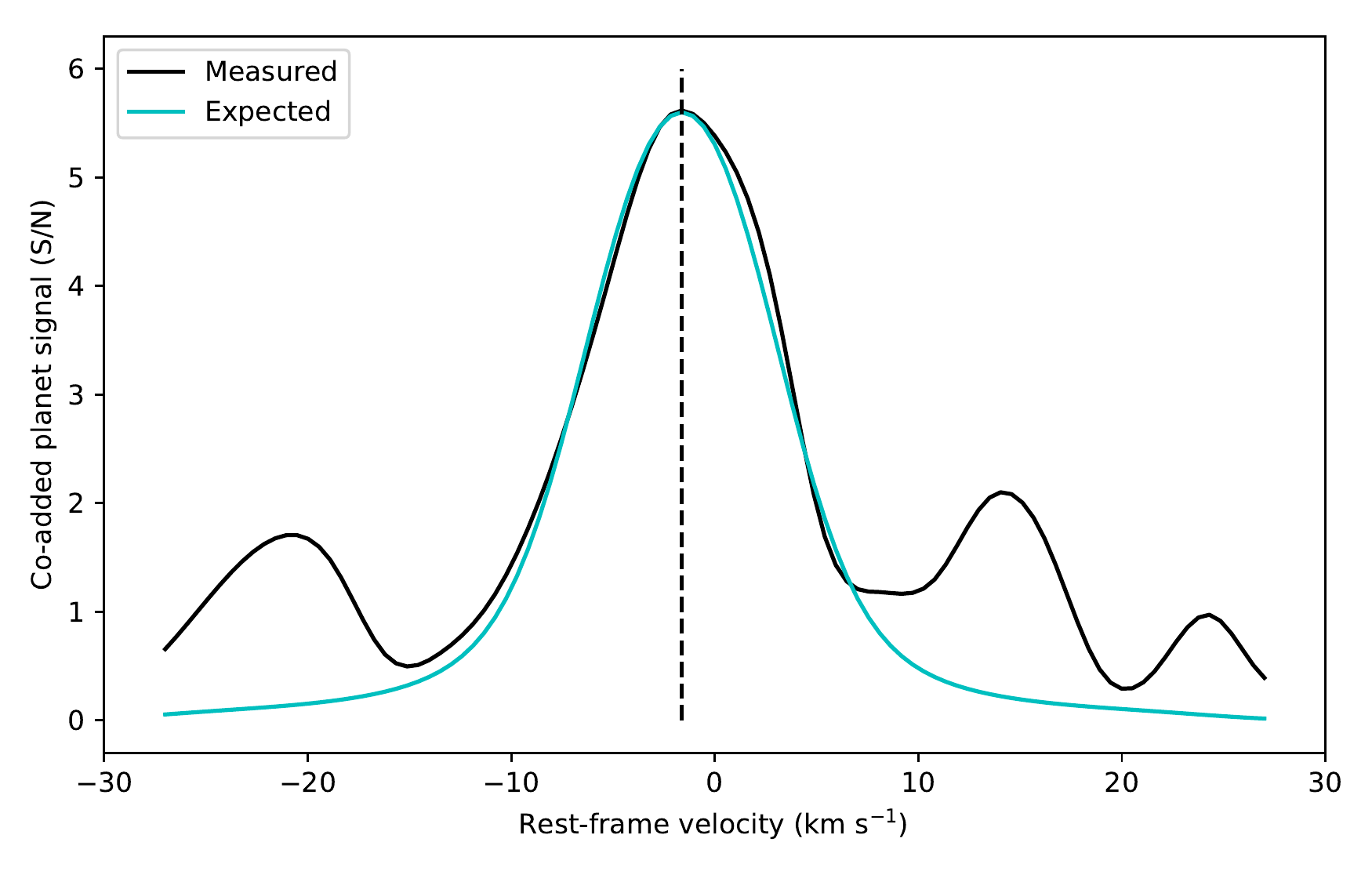}
\caption{Shape of the measured planet signal from the first night of observations (black line) compared to the theoretical analogue (cyan line) computed by convolving the autocorrelation function of the best-fitting model, a Gaussian instrumental profile with FWHM = 6.0 \kms, and a rotational profile (rigid-body rotation, from \citet{bro16}) with equatorial velocity of 3.4 \kms\ and day-to-night wind speed of $-1.6$~\kms. The agreement between the two profiles is remarkable.}
\label{fig:signal_width}
\end{figure}

In Figure~\ref{fig:signal_width} we compare the width of our detection on the first night of observations to a simulated profile. The latter is the convolution of the autocorrelation function of the best-fitting model, a Gaussian with FWHM = 6.0 \kms\ approximating the instrument profile of GIANO, and a rotational profile obtained with the rigid-rotation model of \citet{bro16} for an equatorial rotation of 3.4 \kms\ (i.e. their best-fitting value). The agreement between the measured and the theoretical widths of the signal is excellent. This further suggests that no other spurious source of correlated noise contribute to the data. 
We note that if we increase the level of telluric noise by progressively re-including those discarded orders where the masking of telluric lines did not work, and/or by reducing the threshold of the mask (see details in Section~\ref{sec:tell_corr}), we observe a progressive broadening of the planet signal especially in the red wing, which is coincident with the position of telluric lines (red-shifted) relative to the planet signal on July 11 (see Figure~\ref{fig:ph-rvs}). The correspondence between the expected and the measured width of the signal is hence an additional indicator for an adequate correction of residual telluric signal.

The marginal detection of a global blue shift on these data suggests that the lower spectral resolution of GIANO compared to CRIRES is unlikely to provide meaningful constraints on the planet rotational rate and/or equatorial winds, which were already hard to constrain with CRIRES at a significance of 7.6$\sigma$. We therefore refrain from applying a more sophisticated model for the planet rotation in this study.

\subsection{Residual telluric absorption and alternative analysis techniques}
In Section~\ref{sec:ccorr} we showed that telluric lines need to be masked even after they are de-trended by airmass to achieve a cross-correlation signal free from telluric contamination. We explained that this is likely due to rapid modulations in the flux of telluric lines due to changes in the water vapour content and possibly instrumental profile, which we cannot model with low-order polynomials. However, these variations are largely common-mode between all the spectral channels affected by telluric lines. This opens up to the possibility of using de-trending algorithms such as principal component analysis (PCA) to model and remove these correlated (in time and in wavelength) signals. 

A PCA algorithm was successfully applied by \citet{dek13} to CRIRES data, and more recently \citet{pis16,pis17} have also utilised it - although along the spectral direction rather than the temporal direction - to clean their data from telluric contamination. We have developed our own version of PCA for de-trending GIANO data. As for the standard pipeline, we process each order separately and then co-add their CCFs at a later stage. So far our results have been sub-par compared to the masking technique, mainly due to the fact that the algorithm struggles in isolating telluric residuals from other time-dependent noise components:
\begin{itemize}
\item For an unweighted PCA (i.e. where each spectral channel is equally weighted), the most significant eigenvectors do not contain telluric residuals, but rather trends in spectral channels with very low flux (for instance at the centre of a saturated telluric line) and broad-band variations of the throughput (for instance due to the instrument modal noise).
\item Even when deep absorption lines are masked and the signal-to-noise of each channel is used as weighting, the eigenvectors still model broad-band sources of correlated noise. The main reason for this is that broad-band variations affect a larger portion of the data than narrow telluric lines.
\item If more eigenvectors are linearly combined and removed from the data, telluric residuals are eventually suppressed, but so is the planet signal, resulting in a marginal detection around S/N = 3.
\end{itemize}

\subsection{The non-detection from the second transit}\label{sec:2nd_night}

\begin{figure}[ht!]
\centering
\includegraphics[width=8cm]{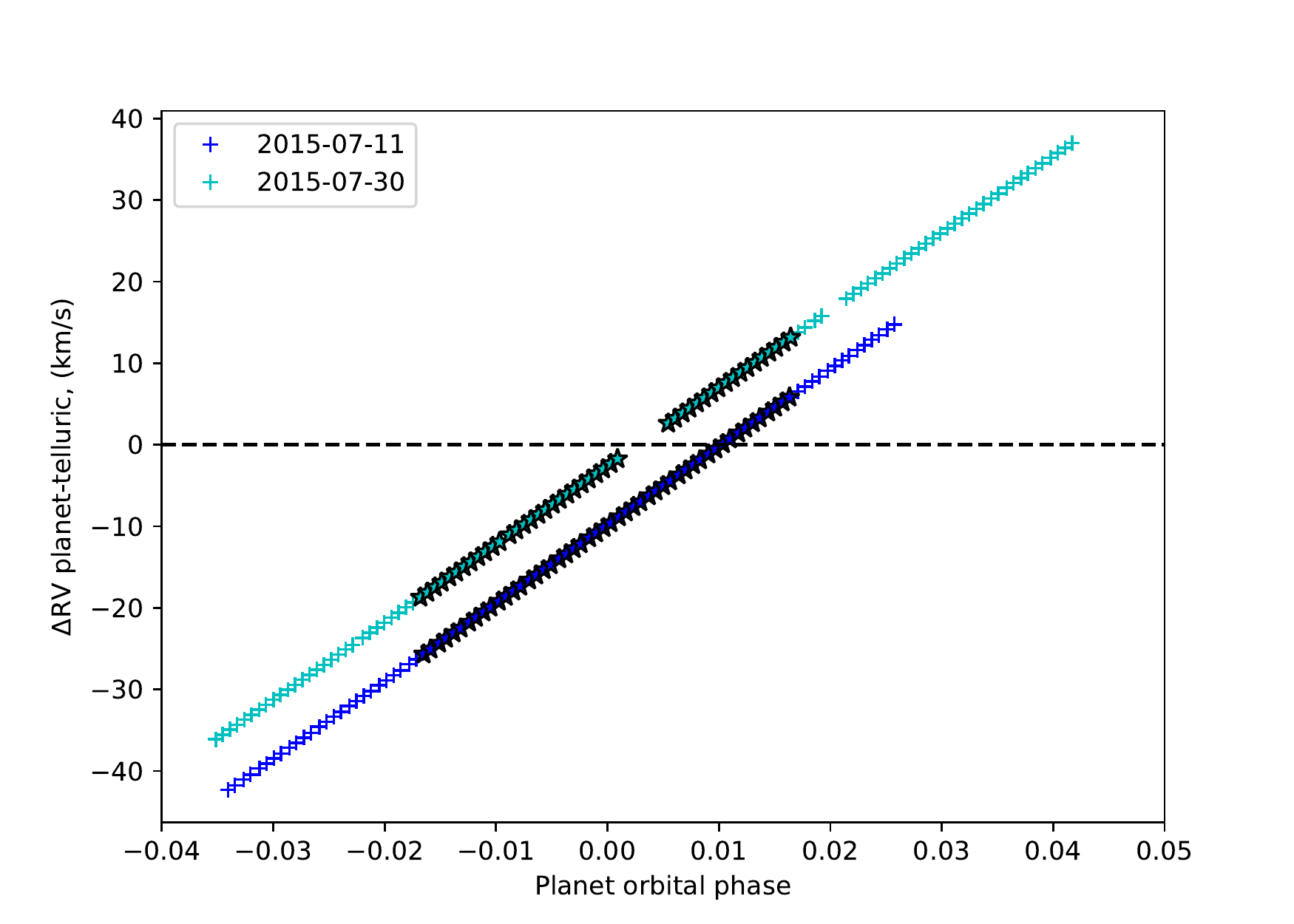}
\caption{Radial velocity of \pname\ as a function of planet orbital phase, relative to the observer. This is the sum of the systemic velocity of the system, and the velocity of the barycentre of the solar system compared to Earth. Symbols and colours are the same as in Figure~\ref{fig:ph-air}. It shows that on the second observing night (light blue) the planet has zero radial velocity close to mid-transit, i.e. when the strength of the transmission spectrum of \pname\ is maximum. We comment on the differences between the two observing nights in Section~\ref{sec:2nd_night}.}
\label{fig:ph-rvs}
\end{figure}

Although the weather and the seeing were good on both nights, several indicators point to an inferior quality of the data taken on July 30. Firstly, the bottom-right panel of Figure~\ref{fig:shifts} shows that the shifts in the dispersion solution as recorded on the GIANO detector are one order of magnitude bigger than on the first night, pointing to a lower stability of the spectrograph. Counter-intuitively, we suspect that this is due to a particularly good seeing. This enhances the non-uniform illumination affecting the fibres of GIANO, and indeed we also measure an enhanced amount of modal noise in the K-band orders (0-7 in our numbering).
Secondly, on July 30 the observations were interrupted twice for technical problems with the instrument and its interface. It resulted in a loss of 20\% of the planetary transit (see gaps in Figures~\ref{fig:ph-air} and \ref{fig:ph-rvs}) and it potentially impacted the overall stability of the instrument.
Finally, Figure~\ref{fig:ph-rvs} shows the radial velocity of \pname\ in the observer's reference frame. This also equals the relative velocity shift between telluric and planet spectra. Ideally one would want to have this shift be as large as possible to minimise the contamination between the two spectra. During the first night of observations (dark blue) the relative radial velocity between \pname\ and the Earth was largely non-zero, except for part of the egress. However, during the second observing night (light blue) the planet radial velocity is zero exactly at mid-transit, which is when the strength of the planet's transmission spectrum is maximum. As a consequence, the masking applied to telluric lines prior to cross-correlation will also potentially affect planet lines, as they are superimposed at mid-transit. The spectra on the second night were also taken at higher airmass on average (Figure~\ref{fig:ph-air}), requiring a more aggressive masking. If this choice succeeded in minimising correlated noise at low \kp\ (Figure~\ref{fig:diags}, bottom panels), it also eroded a larger fraction of the planet spectrum, decreasing our sensitivity.

We substantiate the claim of an inferior data quality on July 30 by injecting the best-fitting planet model in the spectra from both nights, at three times the nominal level, and at the planet radial velocity obtained from Eq.~\ref{eq:pl_vel} with the literature values listed in Section~\ref{sec:ccorr}. After running the injected spectra through the full data-analysis pipeline, we recover a signal-to-noise ratio of 9.4 and 4.3 on the first and second night, respectively. Given that the real planet signal is detected at S/N=4.9 on the first night, scaling arguments point to an expected S/N of 2.2 on the second night, well below detectability. 
Based on the evidence presented in this Section, we conclude that the absence of signal on July 30 is in line with the expectations.

\subsection{Comparison with previous detections}\label{sec:snr}
It is useful to assess whether the detection presented here is compatible with the previous detection of water at 4.8 $\sigma$ with CRIRES \citep{bro16}. For GIANO, we measure 2\,200 $e^-$ per 60 s integration per resolution element (1 pixel = 2.7 \kms) in the K band. With CRIRES, we measure 6\,000 $e^{-}$ per 10 s integration, per resolution element (1 pixel = 1.5 \kms). Scaling the latter signal based on telescope size and pixel width while assuming equal throughput, we would expect from GIANO approximately 12\,500 $e^-$ per 60 s integration per pixel. Observed GIANO spectra are thus deficient by a factor of (12.5 / 2.2) = 5.7, which we attribute to a much lower throughput. Based on flux alone, GIANO should only produce 40\% of the S/N of CRIRES for the same exposure time, insufficient to detect the transmission spectrum of \pname. However, we should also consider the increased spectral range of GIANO. We estimated this additional factor by adding random noise to the best fitting model for the planet transmission spectrum, cutting it accordingly to the wavelength range of GIANO and CRIRES, and cross-correlating it with the noiseless model. For each of the GIANO orders, the noise is rescaled to properly account for the wavelength-dependent transparency of the Earth's atmosphere. We repeat the simulations 100 times, and we compare the distributions of the retrieved S/N for CRIRES and GIANO. On average, GIANO delivers 4.5$\times$ the S/N of CRIRES based on the spectral range alone. However, GIANO has half of the spectral resolution, producing a loss of 30\% in S/N for unresolved planetary lines. Putting all the scaling factors together, \water\ should be detectable with GIANO in one transit at a significance of $4.8\sigma \times 4.5 \times 0.7 \times 0.4 = 6 \sigma$. This is consistent with our reported detection at 5.5 $\sigma$.
Not only does this show that a detection at high significance is possible with just 110 minutes of TNG time, but also that the overall strength of the planet's transmission spectrum has not changed significantly since the previous CRIRES observations from July 2012. Future monitoring of the planet via the same technique will allow us to put constraints on the level of variability of the NIR transmission spectrum due for instance to aerosols at the planet terminator.

\section{Discussion}\label{sec:discussion}
We have tested whether or not a 4-m telescope with a performing high-resolution spectrograph can successfully study the atmospheres of exoplanets at high spectral resolution. We have shown that GIANO in its initial fiber-fed configuration, despite the much smaller throughput, is capable of rivalling CRIRES at the VLT in detecting molecular absorbers with broad-band opacity, such as water vapour. We have demonstrated that the analysis devised for CRIRES still works at half of the spectral resolution, but residual telluric residuals need to be treated more aggressively (e.g. masked) in order to avoid contamination of the planet signal. Smart scheduling of the observations in order to maximise the modulus of the combined barycentric plus systemic velocity of the target is strongly advised to further reduce time-correlated noise coming from uncorrected telluric absorption.

GIANO has been recently updated to a slit-fed instrument and coupled to HARPS-N to deliver quasi-contiguous wavelength coverage from 0.383 to 2.45 \micron\ \citep{cla17}. In this configuration, the instrument will allow us to access in a single observation much smaller atmospheric pressures through sodium absorption \citep{wyt15, lou15}. Furthermore, it will be possible to search for the signature of the weaker absorption bands of water vapour \citep{all17, est17}, which are very sensitive to the presence of clouds or hazes. On the instrumental side, this upgrade enhanced the throughput of GIANO by at least a factor of four. Furthermore, it eliminated the modal noise affecting the orders in the K band (approximately orders 0-7 in our analysis, which we largely discarded). 
When the simulations of Section~\ref{sec:snr} are updated to account for these better performances, and neglecting decreased performances in correcting for telluric lines, we conclude that GIANO could potentially deliver 2.5-3$\times$ the S/N of CRIRES at the VLT in the same observing time, when looking for broad-band absorbers such as \water\ or \methane.This advantage could be used either to open up high-resolution spectroscopy to fainter systems or to start investigating the atmospheres of smaller and cooler exoplanets. 

The planned upgrade for CRIRES \citep[CRIRES+, see][]{criplus} will transform the instrument into a cross-dispersed spectrograph, bringing a 10$\times$ increase in spectral range. Together with state-of-the-art detectors, a refurbished AO system, and improved calibration sources, CRIRES+ is likely to overcome any gaps in sensitivity with GIARPS.  Whereas in the past high-dispersion observations have been limited to either the southern or the northern hemisphere, with both spectrographs online we will be able to access the entire sky.

Adding to the fleet of near-infrared, high-resolution spectrographs, IGRINS \citep{par14} and CARMENES \citep{qui14} have recently come online, and SPIROU \citep{art14} will be available soon. These instruments are revolutionary both in terms of spectral range and throughput compared to the old CRIRES. If the NIR sky at their observatories is of comparable quality to La Palma or Cerro Paranal, and if their initial specifications will be confirmed (especially in terms of throughput), these instruments will be well suited to perform studies of exoplanet atmospheres such as the one presented here.

Importantly for future comparative characterisation of exoplanets, spectra taken at high- and low-resolution contain highly-complementary information \citep{pin17}. Whereas at high resolution temperature and opacity in the planet's atmosphere are encoded in the relative ratio between hundreds of thousands of spectral lines, at low resolution the same information translates into broad-band spectral variations. \citet{bro17} recently demonstrated that constraints on abundances, temperature-pressure ($T$-$p$) profile, and consequently metallicity of exoplanets can be greatly enhanced when combining the confidence intervals from space (HST/WFC3 and Spitzer/IRAC data) and ground observations (CRIRES). To enable this synergy, especially in the era of TESS and JWST, it is necessary to extend high-resolution spectroscopic observations of exoplanets to as many ground-based telescopes as possible. Based on the results of this paper, and considering the increased ease of scheduling, a 3-4-m telescope equipped with a state-of-the art high-resolution spectrograph could not only assist, but even outperform the biggest ground-based telescopes for detecting key molecular constituents such as \water\ and \methane.

The above observations make use of both spectral and temporal resolution to extract exoplanet signals from the overwhelming stellar and telluric contaminants. With a total integration time of ten hours, core-to-wings line ratios as small as 10$^{-5}$ (1$\sigma$) have been detected through their combined cross-correlation signal. 
Contrast ratios exceeding a billion could potentially be reached if the spatial separation is also used, i.e. by coupling a high-resolution spectrograph to a coronographic or extreme-AO system \citep{sne15}. Early demonstrations of the principle were recently presented \citep{sne14, sch16}, and the prospect of targeting biomarkers in terrestrial exoplanets orbiting in the habitable zones of M-dwarfs seems feasible when the next generation of extremely large telescopes come online \citep{sne13, rod14}.

\begin{acknowledgements}
We thank the referee M. Kuerster for his insightful comments, which contributed to improving the quality of the manuscript. We thank R. Claudi and S. Benatti for their insightful discussion about GIARPS. We thank F. Borsa, A. Maggio, N. Nikolov, and I. Pagano for their initial contribution to the design of these high resolution observations. M.B. acknowledges partial support by NASA, through Hubble Fellowship grant HST-HF2-51336 awarded by the Space Telescope Science Institute. P.G. gratefully acknowledges support from INAF through the Progetti Premiali funding scheme of the Italian Ministry of Education, University, and Research. G.G. acknowledges the financial support of the 2017 PhD fellowship programme of INAF. The research leading to these results has received funding from the European Union Seventh Framework Programme (FP7/2007-2013) under Grant Agreement No. 313014 (ETAEARTH).
\end{acknowledgements}

%
%
\bibliographystyle{aa}
\bibliography{gianorefs}

\end{document}